\documentclass[submission,copyright,creativecommons]{eptcs}
\providecommand{\event}{ML 2017}
\usepackage{underscore}          

\usepackage{url}
\usepackage{listings}
\usepackage{color}
\usepackage{abstract}
\usepackage{endnotes}
\usepackage{tikz}
\usepackage{amssymb}
\usepackage{breakurl}
\PassOptionsToPackage{hyphens}{url}
\definecolor{mauve}{rgb}{0.58,0,0.82}
\definecolor{darkpurple}{rgb}{0.3,0,0.3}
\definecolor{darkgreen}{rgb}{0.0,0.3,0}
\definecolor{linkcol}{rgb}{0.0,0.0,0.35}

\usepackage{amsmath,xcolor,listings,url,quoting,multicol}
\usepackage[compact]{titlesec}
\definecolor{darkgreen}{rgb}{0,0.2,0}
\definecolor{darkpurple}{rgb}{0.3,0.1,0.3}

\lstdefinestyle{SML}{
  basicstyle=\ttfamily\normalsize,
  basewidth=0.5em,
  commentstyle=\color{darkgreen},
  escapeinside={(**}{)},
  keywordstyle=\color{darkpurple}\bfseries,
  language=ML,
  stringstyle=\color{blue},
  showstringspaces=false,
  mathescape=true,
  moredelim=**[is][]{?}{?},
  moredelim=**[is][]{&}{&},
  language=ML,
}

\lstdefinelanguage{OCaml}{
  keywords={
    and,as,assert,asr,begin,class,constraint,do,done,downto,effect,else,end,exception,
    external,false,for,fun,function,functor,if,implicit,in,include,inherit,initializer,
    land,lazy,let,lor,lsl,lsr,lxor,macro,match,method,mod,module,mutable,new,nonrec,object,
    of,open,or,private,protected,rec,sig,struct,then,to,true,try,type,val,virtual,when,
    with,while},
  comment=[s]{(*}{*)},
}

\lstdefinestyle{OCaml}{
  basicstyle=\ttfamily\normalsize,
  basewidth=0.5em,
  commentstyle=\color{darkgreen},
  escapeinside={(**}{)},
  keywordstyle=\color{darkpurple}\bfseries,
  language=OCaml,
  stringstyle=\color{blue},
  showstringspaces=false,
  mathescape=true,
  moredelim=**[is][]{?}{?},
  moredelim=**[is][]{&}{&},
}
\title{Extending OCaml's \texttt{open}}
\author{
Runhang Li
\institute{Twitter, Inc}
\email{rli@twitter.com}
\and
Jeremy Yallop
\institute{University of Cambridge}
\email{jeremy.yallop@cl.cam.ac.uk}}

\def\titlerunning{Extending OCaml's \texttt{open}}
\def\authorrunning{R. Li \& J. Yallop}
\begin{document}
\maketitle

\begin{abstract}
We propose a harmonious extension of OCaml's \lstinline{open}
construct.

OCaml's existing construct \lstinline{open M} imports the names
exported by the module \lstinline{M} into the current scope.
At present \lstinline{M} is required to be the \textit{path} to a module.
We propose extending \lstinline{open} to instead accept an arbitrary
module \textit{expression}, making it possible to succinctly address a
number of existing scope-related difficulties that arise when writing
OCaml programs.

\end{abstract}

\section{Introduction: \lstinline[basicstyle=\ttfamily\Large]{open} vs \lstinline[basicstyle=\ttfamily\Large]{include}}

Programming languages intended for large-scale, modular
programming often include features for making names defined in one scope
available without qualification in another scope.
OCaml provides two such operations, via the keywords \lstinline{open}
and \lstinline{include}:
\begin{center}\begin{tabular}{c}\begin{lstlisting}
open M              include M
\end{lstlisting}\end{tabular}\end{center}
Both of these operations introduce the bindings exported by the module
\lstinline{M} into the current scope.
Additionally, \lstinline{include} re-exports the bindings from the
current scope.
This distinction is a useful one, since it is not always appropriate
to re-export the names used within a module.

A second difference between \lstinline{open} and
\lstinline{include} concerns the form of the argument.
In OCaml the argument to \lstinline{open} is a module path:
\begin{center}\begin{tabular}{c}\begin{lstlisting}
open A.B.C
\end{lstlisting}\end{tabular}\end{center}
In contrast, the argument to \lstinline{include} can be any module
expression, such as a functor application, signature-constrained
expression, or structure body:
\begin{center}\begin{tabular}{c}\begin{lstlisting}
include F(X)
include (M:S)
include struct$\ldots$end
\end{lstlisting}\end{tabular}\end{center}
This distinction is less useful: there is no fundamental reason why
\lstinline{include} should accept arbitrary module expressions, while
\lstinline{open} should not.

This paper explores the consequences of extending \lstinline{open}
to eliminate the second difference, so that both \lstinline{open}
and \lstinline{include} accept an arbitrary module expression as
argument (Figure~\ref{figure:extended-open}).
In practice, allowing the form \lstinline{open struct $\ldots$ end}
extends the language with a non-exporting version of every type of
declaration, since any declaration can appear between
\lstinline{struct} and \lstinline{end}.

\begin{figure}[t!]
\begin{minipage}{0.35\textwidth}
\textit{Current design}:\\
Only basic paths are allowed
\begin{center}\lstinline{open M.N}\\~\end{center}
\end{minipage}%
\begin{minipage}{0.1\textwidth}
~
\end{minipage}%
\begin{minipage}{0.55\textwidth}
\textit{Extended design (this paper)}:\\
Arbitrary module expressions are allowed

\begin{center}
\lstinline{open M.N} \quad
\lstinline{open F(M)} \quad
\lstinline{open (M:S)} \quad
\lstinline{open struct $\ldots$ end}
\end{center}
\end{minipage}%

\caption{\label{figure:extended-open}The \lstinline{open} construct and our proposed extension}
\end{figure}

The extended \lstinline{open} has many useful applications, as we illustrate with
examples condensed from
real code (Section~\ref{section:use-cases}).
Our design also resolves some problems in
OCaml's signature language (Section~\ref{section:signature-use-cases}).
We touch briefly
on restrictions and other design considerations (Section~\ref{section:restrictions})
before sketching the implementation (Section~\ref{section:implementation})
and comparing some alternative designs (Section~\ref{section:alternatives}).

\subsection{Status}
\label{section:status}

Following the presentation of this proposal at the OCaml 2017
workshop~\cite{extending-ocamls-open-workshop}, a variant of this design
was discussed at the Caml developers meeting and accepted for
inclusion into OCaml 4.08.
Section~\ref{section:full-status} gives more details.

\section{Extended \lstinline[basicstyle=\ttfamily\Large]{open} in structures: examples}
\label{section:use-cases}

Effectively managing names and scope is a crucial part of structuring
programs.
The examples in this section show how the lack of a facility for local
(non-exporting) declarations can result in awkward structure or
inappropriate scoping in OCaml programs, and further show how these
problems are eliminated by the extended \lstinline{open} construct\footnote{
The reader familiar with Standard ML will recognise that the
\lstinline[style=SML]{local} construct in that language, an
inspiration for this proposal, can also solve the problems described
here.  We return to this point in Section~\ref{section:sml-local}.}.

\subsection{Unexported top-level values}
\label{section:unexported-top-level}

A straightforward use of the extended \lstinline{open} construct is
the introduction of local declarations that are not exported.  In the
code on the left, \lstinline{x} is available in the remainder of the
enclosing module, but it is not exported from the module, as shown in
the inferred signature on the right:\\

\begin{center}
\begin{tabular}{cc}
\begin{lstlisting}
open struct let x = 3 end  $~$
let y = x
\end{lstlisting}&
\begin{lstlisting}
  (* no entry for x *)
  val y : int
\end{lstlisting}%
\end{tabular}
\end{center}

\subsection {A workaround for type shadowing}  
\label{section:type-shadowing}

One common programming pattern is to export a type \lstinline{t} in
each module.
For example, the standard library defines types \lstinline{Float.t},
\lstinline{String.t}, \lstinline{Complex.t}, and many more.
However, this style leads to problems when the definition of one such
\lstinline{t} must refer to another.
For example, in the following code, renaming the \lstinline{s} to
\lstinline{t} requires some care:
\begin{center}\begin{tabular}{c}\begin{lstlisting}
type s = A
module M = struct
  type t = B of s | C
end
\end{lstlisting}\end{tabular}\end{center}
\begin{tikzpicture}[remember picture,overlay]
\begin{scope}[xshift=1.4cm,yshift=3.55cm]
\draw               (7.55,-2.6) rectangle (8.0,-2.15);
\draw[thick,dotted] (5.4,-1.65) rectangle (5.85,-1.2);
\draw [thick,dotted,->] (7.55,-2.15) to [out=100,in=20] (5.85,-1.2) ;
\end{scope}
\begin{scope}[xshift=1.4cm,yshift=0.8cm]
\draw[thick,dotted] (5.75,-3.1) rectangle (6.25,-2.65);
\draw               (7.55,-3.1) rectangle (8.05,-2.65);
\draw [thick,dotted,->] (7.55,-3.1) to [out=-160,in=-20] (6.25,-3.1) ;
\end{scope}
\end{tikzpicture}
Since type definitions are recursive by default, naively renaming
\lstinline{s} to \lstinline{t} in the definition of \lstinline{M.t}
changes the meaning of the definition so that the argument of
\lstinline{B} now refers to the inner \lstinline{t}:
\begin{center}\begin{tabular}{c}\begin{lstlisting}
type t = A
module M = struct
  type t = B of t | C
end
\end{lstlisting}\end{tabular}\end{center}
%
%
The \lstinline{nonrec} keyword,
 added in a recent version of OCaml
   (4.02.2, released June 2015),
 overrides this default,
 making the definition of \lstinline{t} non-recursive,
 and
 restoring the original meaning:
\begin{center}\begin{tabular}{c}\begin{lstlisting}
type t = A
module M = struct
  type nonrec t = B of t | C
end
\end{lstlisting}\end{tabular}\end{center}%

However, in cases where a single type definition must contain both
recursive references and references to another type of the same name,
\lstinline{nonrec} cannot help.
For example, in the following code, \lstinline{t$_1$} and
\lstinline{t$_2$} cannot both be renamed \lstinline{t}, since both
names are used within a single scope, where all
occurrences of \lstinline{t} must refer to the same type:
\begin{center}\begin{tabular}{c}\begin{lstlisting}
type t$_1$ = A
module M = struct
  type t$_2$ = B of t$_2$ * t$_1$ | C
end
\end{lstlisting}\end{tabular}\end{center}

{\begin{tikzpicture}[remember picture,overlay]

\begin{scope}[xshift=0.1cm,yshift=7.3cm]
\draw               (9.05,-2.6) rectangle (9.5,-2.15);
\draw[thick,dotted] (5.4,-1.65) rectangle (5.85,-1.2);
\draw [thick,dotted,->] (9.05,-2.15) to [out=100,in=20] (5.85,-1.2) ;
\end{scope}

\draw (7.9,0.5) rectangle (8.5,1);
\draw[thick,dotted] (5.9,0.5) rectangle (6.5,1);
\draw [thick,dotted,->] (7.9,0.5) to [out=-160,in=-20] (6.5,0.5) ;

\draw (8.9,0.5) rectangle (9.5,1);
\draw[thick,dotted] (5.5,1.5) rectangle (6.1,2);
\draw[thick,dotted,->] (8.9,1) to [out=100,in=20] (6.1,2);

\draw[thick,dotted] (5.75,-3.1) rectangle (6.25,-2.65);
\draw               (7.55,-3.1) rectangle (8.05,-2.65);
\draw               (8.45,-3.1) rectangle (9,   -2.65);

\draw [thick,dotted,->] (7.55,-3.1) to [out=-160,in=-20] (6.25,-3.1) ;

\draw[thick,dotted] (8.25,-2.6) rectangle (8.8,-2.15);
\draw               (9.25,-2.6) rectangle (9.7,-2.15);

\draw [thick,dotted,->] (9,-2.65) to [out=70,in=40] (8.8,-2.15) ;

\draw[thick,dotted] (5.4,-1.65) rectangle (5.85,-1.2);

\draw [thick,dotted,->] (9.25,-2.15) to [out=100,in=20] (5.85,-1.2) ;
\end{tikzpicture}

\noindent
The extended \lstinline{open} construct resolves the difficulty,
making it possible to give an unexported local alias for the outer
\lstinline{t}:
\begin{center}\begin{tabular}{c}\begin{lstlisting}
type t = A
module M = struct
  open struct type t' = t end
  type t = B of t * t' | C
end
\end{lstlisting}\end{tabular}\end{center}

\noindent
Similarly, for GADT-style definitions~\cite{gadts-ocaml} such as the following
\begin{center}\begin{tabular}{c}\begin{lstlisting}
type t = B : t' -> t
       | C : t
\end{lstlisting}\end{tabular}\end{center}
\lstinline{nonrec} can never be used, since every such definition
refers to the definiendum in the return type of each
constructor\footnote{Mantis Issue 6934: \textit{nonrec misbehaves with
    GADTs} \url{https://caml.inria.fr/mantis/view.php?id=6934}}.

\subsection {Local definitions scoped over several functions}
\label{section:local-definitions}

A common pattern involves defining one or more local definitions for
use within one more more exported functions\footnote{
See \lstinline{draw_poly}, \lstinline{draw_poly_line} and
  \lstinline{dodraw} in the OCaml \lstinline{Graphics} module for an
  example.
\url{https://github.com/ocaml/ocaml/blob/4697ca14/otherlibs/graph/graphics.ml}, lines 105--117}.
Typically, the exported functions are defined using tuple pattern
matching.  Here is an example, defining \lstinline{f} and
\lstinline{g}  in terms of an auxiliary unexported function,
\lstinline{aux}:
\begin{center}\begin{tabular}{c}\begin{lstlisting}
let f, g =
  let aux x y =
     $\ldots$
  in (fun p -> aux p true),
     (fun p -> aux p false)
\end{lstlisting}\end{tabular}\end{center}

\noindent
This style has several drawbacks.
First, the names \lstinline{f} and \lstinline{g} are separated from
their definitions by the definition of \lstinline{aux}.
Second, the unsugared syntax for creating functions \lstinline{fun x -> $\ldots$} must be used in place of the more typical sugared
syntax \lstinline{let f x = $\ldots$}.
Finally, the definition allocates an intermediate tuple.
With the extended \lstinline{open} construct, all of these these
problems disappear:
\begin{center}\begin{tabular}{c}\begin{lstlisting}
include struct
  open struct let aux x y = $\ldots$ end
  let f p = aux p true
  let g p = aux p false
end
\end{lstlisting}\end{tabular}\end{center}

\noindent
The surrounding \lstinline{include struct $\ldots$ end} delimits the
scope of the local binding \lstinline{aux}, so that \lstinline{aux} is only visible
in the definitions of \lstinline{f} and \lstinline{g}, not in the code
that follows.

\subsection {Local exception definitions}
\label{section:local-exception-definitions}

OCaml's \lstinline{let module} construct supports defining exceptions
whose names are visible only within a particular expression.
For example, in the following code, the \lstinline{Interrupt}
exception is only visible within the body of the
\lstinline{let module $\ldots$ in}
binding:
\begin{center}\begin{tabular}{c}\begin{lstlisting}
let module M = struct exception Interrupt end in
  let rec loop () = ... raise M.Interrupt
      and run () = match loop () with
    | exception M.Interrupt -> Error "failed"
    | x -> Ok x
  in run ()
\end{lstlisting}\end{tabular}\end{center}
Since OCaml 4.04, a construct that supports local exceptions more directly
is also available~\cite{gpr-638}}:
\begin{center}\begin{tabular}{c}\begin{lstlisting}
let exception Interrupt in
  let rec loop () = ... raise Interrupt
      and run () = match loop () with
    | exception Interrupt -> Error "failed"
    | x -> Ok x
  in run ()
\end{lstlisting}\end{tabular}\end{center}
Limiting the scope of exceptions supports a common idiom
in which exceptions are used to pass information between a raiser and
a handler without the possibility of
interception~\cite{exceptions-are-shared-secrets}.  (This idiom is
perhaps even more useful for programming with
effects~\cite{ocaml-effects}, where information flows in both
directions.)

Limiting the scope of exceptions can make control flow easier to
understand and, in principle, easier to optimize; in some cases,
locally-scoped exceptions can be compiled using local
jumps~\cite{gpr-638}.

The extended \lstinline{open} construct improves support for this
pattern.  While \lstinline{let module} allows defining exceptions
whose names are visible only within particular expressions, the extended
\lstinline{open} also allows limiting visibility to particular
declarations.
In the following snippet, the \lstinline{Interrupt} exception is only
visible in the definitions of \lstinline{loop} and \lstinline{run}:
\begin{center}\begin{tabular}{c}\begin{lstlisting}
include struct
  open struct exception Interrupt end
  let rec loop () = ... raise Interrupt
      and run () = match loop () with
    | exception Interrupt -> Error "failed"
    | x -> Ok x
end
\end{lstlisting}\end{tabular}\end{center}

As with the previous example, this style of local definition is
supported in Standard ML by the \lstinline[style=SML]{local} construct
discussed in Section~\ref{section:sml-local}.

\subsection {Shared state}
\label{section:shared-state}

Similarly, the extended \lstinline{open} supports
limiting the scope of global state to a particular set of declarations: 
\begin{center}\begin{tabular}{c}\begin{lstlisting}
open struct
  open struct let counter = ref 0 end
  let inc () = incr counter
  let dec () = decr counter
  let current () = !counter
end
\end{lstlisting}\end{tabular}\end{center}

Here the names \lstinline{inc}, \lstinline{dec} and
\lstinline{current} are accessible in the code that follows, but the
shared reference \lstinline{counter} is not.

\subsection{Local names in generated code}
\label{section:generated-names}

It is common in OCaml to use low-level code generation in the
implementation of libraries and programs.

Until recently, the most common system for compile-time code
generation was the Camlp4 preprocessor that performs transformations on the concrete
syntax of programs.
These transformations can result in the generation of entirely new functions and
modules as is the case with the \texttt{deriving} framework that generates
pretty-printers, serializers, and other functions from type
definitions~\cite{deriving}.

More recently, the \texttt{ppx} framework, which supports
transformations on abstract syntax~\cite{ppx}, has become popular.
Syntax transformers based on \texttt{ppx}, such as
\texttt{ppx\_deriving}
(a reimplementation of the \texttt{deriving} generic programming framework~\cite{deriving}),
\texttt{js\_of\_ocaml-ppx}
(an extension for manipulating JavaScript properties, distributed as part of the \texttt{js\_of\_ocaml} OCaml-to-JavaScript compiler~\cite{js-of-ocaml}),
\texttt{ppx\_lwt} (a syntax for constructing promise computations, part of the \texttt{lwt} lightweight concurrency
framework~\cite{lwt}) and
\texttt{ppx\_stage} (a preprocessor for typed multi-stage
programming),
may also generate large amounts of code.

The definitions introduced by Camlp4 and \texttt{ppx} extensions are
often intended to be details of the implementation, not exposed to
the programmer, and with names that do not interact with the remainder
of the program.
However, it is currently difficult to introduce completely anonymous
declarations in OCaml.
A common solution is to generate instead a module with a
``sufficiently unique'' name --- i.e.~a name that is unlikely to clash
with names defined by the programmer.
For example, here is a simple expression, representing a function that
generates a code fragment, written using \texttt{ppx\_stage}:

\begin{center}\begin{tabular}{c}\begin{lstlisting}
fun x -> [%code [%e x] ]
\end{lstlisting}\end{tabular}\end{center}

The \texttt{ppx\_stage} extension transforms the function body to
generate a module with various components that implement the behaviour
of the code fragment:

\begin{center}\begin{tabular}{c}\begin{lstlisting}
module Staged_349289618 =
struct
  let staged0 hole''_1 =
  let contents''_1 = hole''_1  in
   ...
\end{lstlisting}\end{tabular}\end{center}

If, as is often the case, the user of \texttt{ppx\_stage} does not
provide an interface file, the generated module
\texttt{Staged\_349289618} will appear in the interface to the module,
exposing the internal details of the code generation scheme.

\subsection{Restricted open}
\label{section:restricted-open}

It is sometimes useful to import a module
under a restricted signature.  For example, the following statement

\begin{center}\begin{tabular}{c}\begin{lstlisting}
open (Option : MONAD)
\end{lstlisting}\end{tabular}\end{center}

\noindent
imports only those identifiers from
the \lstinline{Option} module that appear in the \lstinline{MONAD}
signature.

There is a caveat here: besides excluding identifiers not
found in \lstinline{MONAD}, OCaml's module ascription also hides
concrete type definitions behind abstract types, which is typically
not the desired behaviour for \lstinline{open}.
This behaviour can be avoided by adding an explicit constraint to the
constraining \lstinline{MONAD} signature to maintain the equality
between the type \lstinline{t} in the signature and
\lstinline{Option.t}:

\begin{center}\begin{tabular}{c}\begin{lstlisting}
open (Option : MONAD with type 'a t = 'a Option.t)
\end{lstlisting}\end{tabular}\end{center}

However, this is rather verbose.  The difficulty could be more
succinctly addressed by extending OCaml with a construct found in
Standard ML, namely transparent signature
ascription~\cite{harper-lillibridge}, a useful feature in its own
right.

\section{Extended \lstinline[basicstyle=\ttfamily\Large]{open} in signatures: examples}
\label{section:signature-use-cases}

In signatures, as in structures, the argument of \lstinline{open} is
currently restricted to a qualified module path
(Figure~\ref{figure:extended-open}).
As in structures, we propose extending \lstinline{open} in signatures
to allow an arbitrary module expression as argument.
However, while extended \lstinline{open} in structures evaluates its
argument, \lstinline{open} in signatures is used only during type
checking.

This section presents examples of signatures that benefit from the
extended \lstinline{open}.
Our examples all involve type definitions, but it is possible to
construct similar examples for other language constructs, such as
functors and classes.

\subsection{Unwriteable, unprintable signatures}
\label{section:unprintable-signatures}

The OCaml compiler has a feature that is often useful during
development: passing the \verb|-i| flag when compiling a module causes
OCaml to display the inferred signature of the module.
However, users are sometimes surprised to find that a signature
generated by OCaml is subsequently rejected by OCaml, because it is
incompatible with the original module, or even because it is invalid
when considered in isolation.

Here is an example of the first case.  The signature on the right is
the output of \verb|ocamlc -i| for the module on the left:

\begin{center}
\begin{tabular}{cp{0.5in}c}
\begin{lstlisting}
type t = T1
module M = struct
  type t = T2
  let f T1 = T2
end
\end{lstlisting}&&
\begin{lstlisting}
type t = T1
module M : sig
  type t = T2
  val f : t -> t
end
\end{lstlisting}
\end{tabular}
\end{center}

\noindent
The input and output types of \lstinline{M.f} are different in the
module, but printed identically.
That is, the printed type for \lstinline{f} is incorrect.

Here is an example of the second case, again with the original module
on the left and the generated signature on the right:

\begin{center}
\begin{tabular}{cp{0.5in}c}
\begin{lstlisting}
type t = T
module M = struct
  type 'a t = S
  let f T = S
end
\end{lstlisting}&&
\begin{lstlisting}
type t = T
module M : sig
  type 'a t = S
  val f : t -> t
 end
\end{lstlisting}
\end{tabular}
\end{center}

This time the generated signature is ill-formed because the type
\lstinline{M.t} requires a type argument, but is used without one.

If these problems arose from a shortcoming in the implementation of
the \verb|-i| flag then there would be little cause for
concern.
In fact, they point to a more fundamental issue: many OCaml
modules have signatures that cannot be given a printed representation.
It is impossible to generate suitable signatures; more
importantly, it is impossible even to write down suitable signatures
by hand.

The problem in both cases is scoping: an identifier such as
\lstinline{t} always refers to the most recent definition, and there
is no way to refer to other bindings for the same name.
The \lstinline{nonrec} keyword (Section~\ref{section:type-shadowing}), solves a
few special cases of the problem, by making it possible to refer to a
single other definition for \lstinline{t} within the definition of
\lstinline{t} itself.
But most such problems, including the examples above, are not solved
by \lstinline{nonrec}.

The extended \lstinline{open} solves the problem entirely,
by making it possible to give internal aliases to names.
For example, here is a valid signature for the first case above using
the extended \lstinline{open}.

\begin{center}
\begin{tabular}{cc}
\begin{lstlisting}
type t = T1
module M = struct
  type t = T2
  let f T1 = T2
end
\end{lstlisting}&
\begin{lstlisting}
type t = T1
open struct type t' = t end
module M : sig
  type t = T2
  val f : t' -> t
end
\end{lstlisting}
\end{tabular}
\end{center}

The OCaml compiler might similarly insert a minimal set of aliases to
resolve shadowing without the need for user intervention.  (At the
time of writing, however, our implementation does not yet include this
improvement to signature printing.)

And, of course, the extended \lstinline{open} also makes it possible for
users to write those signatures that are currently inexpressible.


\subsection{Local type aliases in signatures}
\label{section:local-type-aliases}

Even in cases with no shadowing, it is sometimes useful to define a
local type alias in a signature\footnote{
For example, the functions
\lstinline{comment},
\lstinline{maintainer},
 \lstinline{run},
 \lstinline{cmd},
 \lstinline{user},
\lstinline{workdir},
 \lstinline{volume}, and
 \lstinline{entrypoint} in the \texttt{Dockerfile} module would benefit from such an alias.
 \url{https://github.com/avsm/ocaml-dockerfile/blob/e0dad1a/src/dockerfile.mli}
 }. In the following code, the type
\lstinline{t} is available for use in \lstinline{x} and \lstinline{y},
but not exported from the signature.

\begin{center}\begin{tabular}{c}\begin{lstlisting}
open struct type t = int -> int end
val x : t
val y : t
\end{lstlisting}\end{tabular}\end{center}



\section{Restrictions and design considerations}
\label{section:restrictions}

\subsection{Dependency elimination}
\label{section:dependency-elimination}

OCaml's applicative functors impose a number of restrictions on
programs beyond type compatibility.
One such restriction arises in functor application: it must be
possible
to ``eliminate'' in the functor result type each type defined in
the functor argument~\cite{Leroy-modular-modules}.
For example, given the following functor definition

\begin{center}\begin{tabular}{c}\begin{lstlisting}
module F(X: sig type t val x: t end) =
struct
 let x = X.x
end
\end{lstlisting}\end{tabular}\end{center}

\noindent
the following application is valid:

\begin{center}\begin{tabular}{c}\begin{lstlisting}
module A = struct type t = T let x = T end
module B = F(A)
\end{lstlisting}\end{tabular}\end{center}

\noindent
and \lstinline{B} receives the following type:

\begin{center}\begin{tabular}{c}\begin{lstlisting}
module B : sig val x : A.t end
\end{lstlisting}\end{tabular}\end{center}

\noindent
However, the following application is not allowed:

\begin{center}\begin{tabular}{c}\begin{lstlisting}
F(struct type t = T let x = T end)
\end{lstlisting}\end{tabular}\end{center}

\noindent
since the result of the application cannot be given a type, as there
is no suitable name for the type of \lstinline{x}.

The extended \lstinline{open} construct has a similar restriction.
For example, the following program is rejected by the
type-checker because the only suitable name for the type of \lstinline{x},
namely \lstinline{t}, is not exported:

\begin{center}\begin{tabular}{c}\begin{lstlisting}
open struct type t = T end
let x = T
\end{lstlisting}\end{tabular}\end{center}

\noindent
Here is the error message from the compiler:

\begin{center}
\begin{minipage}{0.8\textwidth}
\begin{small}
\begin{verbatim}
1 | open struct type t = T end
    ^^^^^^^^^^^^^^^^^^^^^^^^^^
Error: The type t/89 introduced by this open appears in the signature
       Line 2, characters 4-5:
         The value x has no valid type if t/89 is hidden
\end{verbatim}
\end{small}
\end{minipage}
\end{center}

Since the restriction for the extended \lstinline{open} construct is
the same as the existing functor restriction, we can reuse the
existing implementation of the check in the OCaml type checker.
In particular we use the \lstinline{Mtype.nondep_supertype} function
to check if introduced identifiers can be eliminated from rest of the
structure~\cite{Leroy-modular-modules}.

\subsection{The Avoidance Problem}
\label{section:avoidance-problem}

The \textit{avoidance problem}~\cite{dreyer-phd} is closely connected
with dependency elimination.
The problem is as follows: it is sometimes necessary to find a
signature for a module that avoids mention of one of its dependencies;
however, it is not always possible to find a best, or
\textit{principal} (i.e.~most-specific) such signature, since the
candidates may be incomparable.

Dreyer~\cite{dreyer-phd} gives the following example of the surprising
behaviour that can arise from OCaml's lack of principal signatures.
Suppose a signature \lstinline{S}, and two functors \lstinline{F} and
\lstinline{G} that each take an argument of type \lstinline{S}, as
follows:

\begin{center}\begin{tabular}{c}\begin{lstlisting}
module type S = sig type t end
module F (X : S) = struct type u = X.t type v = X.t end
module G (X : S) = struct type u = X.t type v = u end
\end{lstlisting}\end{tabular}\end{center}

\noindent
Semantically, \lstinline{F} and \lstinline{G} are equivalent: in both
cases, the types \lstinline{u}, \lstinline{v} and \lstinline{X.t} are
all equal in the body of the functor.
If \lstinline{F} and \lstinline{G} are applied to a module denoted by
a path, then the resulting signatures are equivalent.
For example, here is the result of applying \lstinline{F} and
\lstinline{G} to the top-level module \lstinline{Char}:

\begin{center}\begin{tabular}{c}\begin{lstlisting}
# module FC = F(Char);;
module FC : sig type u = Char.t type v = Char.t end
# module GC = G(Char);;
module GC : sig type u = Char.t type v = u end
\end{lstlisting}\end{tabular}\end{center}

\noindent
Since the argument \lstinline{Char} has a globally-visible name, OCaml
is able to preserve all the equalities in the output types.

However, when the module passed as argument is not denoted by a path
then the result of applying \lstinline{F} is different from the result
of applying \lstinline{G}~\cite{leroy-applicative}:

\begin{center}\begin{tabular}{c}\begin{lstlisting}
# module FI = F(struct type t = int end : S);;
module FI : sig type u type v end
# module GI = G(struct type t = int end : S);;
module GI : sig type u type v = u end
\end{lstlisting}\end{tabular}\end{center}

\noindent
This time OCaml cannot preserve all the equalities, since there is no
way of naming the type member of the module passed as argument in the
output signature.
Consequently, the type equalities that syntactically involve
\lstinline{X.t} are discarded, making the types \lstinline{FI.u}
\lstinline{FI.v}, and \lstinline{GI.u} abstract.

A similar situation arises with the extended \lstinline{open}
construct, which inherits OCaml's approach towards elimination of
modules in signatures.

In the following examples \lstinline{M} is given a less general type
than \lstinline{N}, even though the two modules are semantically
equivalent:

\begin{center}\begin{tabular}{cp{0.2in}c}\begin{lstlisting}
module M = struct
  open struct type t = T end
  type u = t and v = t
end
\end{lstlisting}&&
\begin{lstlisting}
module N = struct
  open struct type t = T end
  type u = t and v = u
end
\end{lstlisting}\end{tabular}\end{center}

\noindent
Here are the types assigned by OCaml:

\begin{center}\begin{tabular}{cp{0.2in}c}\begin{lstlisting}
module M : sig
  type u and v      $~$
end
\end{lstlisting}&&
\begin{lstlisting}
module N : sig
  type u and v = u      $~$
end
\end{lstlisting}\end{tabular}\end{center}

\noindent
As with \lstinline{F} and \lstinline{G}, the type equalities
syntactically involving \lstinline{t} are discarded, even though the
two modules are semantically equivalent, since the types
\lstinline{u}, \lstinline{v} and \lstinline{t} are all equal in each case.

\subsection{Evaluation of extended \lstinline{open} in signatures}
\label{section:evaluation-signatures}

Here is a possible objection to supporting the extended
\lstinline{open} in signatures: although local type definitions are
useful within signatures, local value definitions are not, and so it
would be better to restrict the argument of \lstinline{open} to permit
only type definitions.

For example, the following runs without raising an exception:
\begin{center}\begin{tabular}{c}\begin{lstlisting}
module type S =
sig
  (* no exception! *)
  open struct assert false end
end
\end{lstlisting}\end{tabular}\end{center}
Within a signature, \lstinline{open}'s argument is used only for its
type, and so the expression \lstinline{assert false} is not evaluated.

In fact, this behaviour follows an existing principle of OCaml's
design: \textit{module expressions in type contexts are not
  evaluated}.
For example, the \lstinline{module type of} construct, currently
supported in OCaml, also accepts a module expression that is not
evaluated:
\begin{center}\begin{tabular}{c}\begin{lstlisting}
module type S = (* no exception! *)
  module type of struct assert false end
\end{lstlisting}\end{tabular}\end{center}
And similarly, functor applications that occur within type expressions
in OCaml are not evaluated:
\begin{center}\begin{tabular}{c}\begin{lstlisting}
module F(X: sig end) =
struct
  assert false
  type t = int
end
let f (x: F(List).t) = x (* no exception! *)
\end{lstlisting}\end{tabular}\end{center}

\section{Implementation sketch}
\label{section:implementation}

As the discussion in Sections~\ref{section:avoidance-problem}
and~\ref{section:dependency-elimination} indicates, the subtleties in
the static semantics of the extended \lstinline{open} also occur with
OCaml's functors.
Our implementation takes advantage of this fact, reusing existing
functions in OCaml's type checker.
In particular, the function \lstinline{nondep_supertype}
\begin{center}\begin{tabular}{c}\begin{lstlisting}
val nondep_supertype: Env.t -> Ident.t -> module_type -> module_type
\end{lstlisting}\end{tabular}\end{center}
is used in the
OCaml type checker to eliminate identifiers without paths from the module
types that arise from functor applications; we use it a second time to
eliminate identifiers without paths from the types of the declarations
that follow an occurrence of the extended \lstinline{open}
(Section~\ref{section:dependency-elimination}).
The interested reader may find a fuller description of
\lstinline{nondep_supertype} in Leroy's article on implementing module
systems~\cite{Leroy-modular-modules}.

In more detail, the updated type-checker in our implementation behaves
as follows on encountering the phrase \lstinline{open modexp; decl}.
First, \lstinline{modexp} is type-checked using the function
\lstinline{type_open}, which returns several components:
a fresh name for the module of a form that cannot occur in programs (\lstinline{M#1}, say),
a representation of the module type,
and a corresponding typing environment.
Next, \lstinline{decl} is type-checked in this extended typing environment.
Finally, the type-checking procedure constructs a representation of the extended 
type-checked program \lstinline{module M#1 = modexp; open M#1; decl}.
This representation is ultimately used to generate code: OCaml's
compiler gives modules a run-time representation and an entry in the
parent module; this compilation scheme requires that
\lstinline{modexp} has such a representation, too.

Following this step, the \lstinline{nondep_supertype} function
attempts to eliminate the generated identifier \lstinline{M#1} from
the type of \lstinline{decl}, failing with a user-facing diagnostic if
it cannot be eliminated.
Finally, the entry for \lstinline{M#1} is removed from the type of the
enclosing module, so that it does not appear in types seen by the
user.

The sketch above covers the essence of the implementation.
The full patch also supports local open in signatures
(Section~\ref{section:signature-use-cases}),
let bindings, and signatures.
The interested reader may find the full details in the GitHub pull
request: \url{https://github.com/ocaml/ocaml/pull/1506}.

\section{Alternative designs}
\label{section:alternatives}

The facilities provided by the extended \lstinline{open} are
frequently useful, as the examples in Sections~\ref{section:use-cases}
and~\ref{section:signature-use-cases} indicate, and so it is no
surprise that other languages provide comparable facilities.
This section compares two of these alternatives, based on the keywords
\lstinline[style=SML]{local} and
\lstinline{private}.

\subsection{\lstinline[style=SML,basicstyle=\ttfamily\Large]{local}}
\label{section:sml-local}

The design in this paper draws inspiration from Standard ML's
\lstinline[style=SML]{local} construct~\cite{sml97}:
\begin{center}\begin{tabular}{c}\begin{lstlisting}[style=SML]
local declarations$_1$
   in declarations$_2$
end
\end{lstlisting}\end{tabular}\end{center}

\noindent
As the keyword suggests, names introduced by the first set of
declarations (\lstinline{declarations$_1$}) are in scope only within
the second set \lstinline{declarations$_2$}, not in the code that
follows.

The original 1990 Definition of Standard ML~\cite{sml90} also allows
\lstinline[style=SML]{local} in specifications (signatures), making it
possible to similarly encode the examples of
Section~\ref{section:signature-use-cases}.
The language defined in the 1997 revision of the
Definition~\cite{sml97} no longer allows 
\lstinline[style=SML]{local} in specifications.
However, they are still supported in the latest release of at least
one implementation, Moscow ML~\cite{moscowml}.

To a first approximation\footnote{
There are some inessential differences:
with Standard ML's \lstinline[style=SML]{local},
 type names in \lstinline!declarations!$_1$
   that cannot be eliminated
     in the types of \lstinline!declarations!$_2$ become abstract,
while the corresponding situation with \lstinline{open} is treated as an error
in our proposal
(Section~\ref{section:dependency-elimination}).
},
the \lstinline[style=SML]{local} construct
can be defined straightforwardly in terms of \lstinline{open} as follows:

\begin{center}\begin{tabular}{rcl}
{\lstinline[style=SML]!local d1 in d2 end!}
&$\leadsto$&
{\lstinline!include open struct d1 end d2 end!}\\
\end{tabular}\end{center}

\noindent
The definition of the extended \lstinline{open} in terms of
\lstinline{local} is slightly less straightforward:

\begin{center}\begin{tabular}{rcl}
{\lstinline!open modexp; d!}
&$\leadsto$&
{\lstinline[style=SML]!local structure M = modexp; open M in d end!}\\
&& \qquad (where \lstinline!M! is not free in \lstinline!d!)
\end{tabular}\end{center}

Unlike the translation from \lstinline[style=SML]{local} to
\lstinline{open}, this second translation makes use of the surrounding
context of the translated expression.
First, the declarations \lstinline{d} following the \lstinline{open}
statement are included on the left hand side of the translation;
this makes it possible to delimit the scope of the identifiers imported from
\lstinline{modexp}.
Second, and more significantly, the side condition requires that the name
\lstinline{M} introduced on the right hand side of the translation
does not appear free in \lstinline{d}, to avoid shadowing definitions
in the surrounding context.
In other words, while \lstinline[style=SML]{local} is \textit{macro
  expressible}~\cite{expressive-power} in terms of \lstinline{open},
\lstinline{open} is not macro expressible in terms of
\lstinline[style=SML]{local}.

The reader may note the similarity between the translation of
\lstinline{open} into \lstinline[style=SML]{local} and the elaboration
into a program with a freshly generated module name that occurs during
type-checking of \lstinline{open}  (Section~\ref{section:implementation}).
This generativity appears to be an essential part of the
expressiveness enabled by the extended \lstinline{open}. Unless the type
checker is extended to generate fresh names (as in our
implementation), the expressive power can only be recovered if an
equivalent step is performed by the user (as with the free-variable
check with the translation into local).

The translations show, then, that \lstinline{open} is a little more expressive than 
\lstinline[style=SML]{local}.  In fact, the extra expressiveness is sometimes useful in practice.
Programs that generate code must be careful to avoid name shadowing
(Section~\ref{section:generated-names}).
In OCaml, such programs are typically written as transformations on
untyped abstract syntax trees, for which it is often not possible to
determine whether a variable is free\footnote{ For example, in the
  expression \lstinline!let open M in x + y!  whether \lstinline!x!
  and \lstinline!y! are free depends on whether \lstinline!M! exports
  those identifiers --- that is, it depends upon the type of
  \lstinline!M!.
}.
For such use cases, extended \lstinline{open} is a little more
convenient than \lstinline[style=SML]!local!.



\subsection{\lstinline[basicstyle=\ttfamily\Large]{private}}
\label{section:private}

Many object-oriented languages use a \lstinline{private} keyword
to mark non-exporting declarations.
Indeed, the object oriented part of early versions of OCaml supported
\lstinline{private} instance variables in classes with this meaning:
\begin{center}\begin{tabular}{c}\begin{lstlisting}
class c = object val private x = 3 end
\end{lstlisting}\end{tabular}\end{center}
However, for the last two decades\footnote{
  The following updates to the OCaml compiler and manual removed
  private instance variables
  and introduced private methods with the current semantics:
  \begin{itemize}
  \item J\'er\^ome Vouillon (June 24, 1998): \textit{Nouvelle syntaxe des classes}\\
  \url{https://github.com/ocaml/ocaml/commit/87b17301}
  \item J\'er\^ome Vouillon (August 13, 1998):\textit{Mise a jour des classes},\\
    \url{https://github.com/ocaml/ocaml-manual/commit/63bea030}
  \end{itemize}
}
only private methods, not private instance variables are supported,
and \lstinline{private} has a meaning closer to \lstinline{protected} in other
object-oriented languages, limiting scope to the current class and its sub-classes.

As with \lstinline[style=SML]{local}, it would be possible to support
the examples in Sections~\ref{section:use-cases}
and~\ref{section:signature-use-cases} by adding support for
\lstinline{private} annotations on declarations in structures and
signatures.
However, supporting \lstinline!private! annotations introduces
additional syntactic considerations. 
In particular, it is natural to extend
\lstinline{open} to allow arbitrary module expressions
(since every
form of module expression --- functor application, unnamed structure,
ascription, etc. --- is potentially useful as the argument to
\lstinline{open}),
but it is less natural to support \lstinline{private} annotations
on every type of declaration.
For example, while private type aliases in signatures are clearly
useful (Section~\ref{section:local-type-aliases}), there do not appear
to be any uses for private exception declarations in signatures.
A design based around \lstinline{private} therefore appears to bring a
choice between a uniform but loose grammar (i.e.~with support for
various useless constructs), or a complicated grammar that allows
\lstinline{private} only for constructs where it is useful.

As with \lstinline[style=SML]!local!, it is possible to
define \lstinline{private} in terms of the extended \lstinline{open}:

\begin{center}\begin{tabular}{rcl}
\begin{lstlisting}
private decl
\end{lstlisting}
&$\leadsto$&
\begin{lstlisting}
open struct decl end
\end{lstlisting}
\end{tabular}\end{center}

\noindent
Once again, the definition of \lstinline{open} in terms of
\lstinline{private} is a little less straightforward:

\begin{center}\begin{tabular}{rcl}
{\lstinline!open modexp; decl!}
&$\leadsto$&
{\lstinline!private module M = modexp; open M; decl!}\\
&& \qquad (where \lstinline!M! is not free in \lstinline!decl!)
\end{tabular}\end{center}

\noindent
And, as with the translation from \lstinline{open} into
\lstinline[style=SML]!local!, the translation from \lstinline{open}
into \lstinline{private} involves determining the set of free
identifiers in the declarations that follow, making
\lstinline{private} a less suitable basis than \lstinline{open} for
code generation involving non-exporting declarations
(Section~\ref{section:generated-names}).


\subsection{Signature-local bindings}
\label{section:signature-local-bindings}

An alternative approach to supporting the use cases of
Section~\ref{section:signature-use-cases}
builds on another OCaml feature, \emph{destructive
  substitution}~\cite{destructive-substitution}.
Destructive substitution is an operation on signatures that
simultaneously eliminates a type (or module) component within a
signature and replaces each use of the component with a compatible
type (or module) expression.

For example, here is a definition for a module type \lstinline{T} that
defines a type component \lstinline{t} and a value component
\lstinline{f} whose type uses \lstinline{t}:

\begin{center}\begin{tabular}{c}\begin{lstlisting}
module type T = sig type t val f : t -> t end 
\end{lstlisting}\end{tabular}\end{center}

\noindent
The following code defines a module type \lstinline{S} by eliminating
\lstinline{t} and replacing each occurrence of \lstinline{t} with
\lstinline{int} in the remainder of the module:

\begin{center}\begin{tabular}{c}\begin{lstlisting}
module type S = T with type t := int
\end{lstlisting}\end{tabular}\end{center}

\noindent
The result of this destructive substitution is equivalent to the
following direct definition of \lstinline{S}:

\begin{center}\begin{tabular}{c}\begin{lstlisting}
module type S = sig val f : int -> int end 
\end{lstlisting}\end{tabular}\end{center}

Jacques Garrigue has proposed extending destructive substitution to
local aliases in signatures, so that a definition of the following
form
 
\begin{center}\begin{tabular}{c}\begin{lstlisting}
type t := e
\end{lstlisting}\end{tabular}\end{center}

\noindent
in a signature would behave equivalently to the following extended \lstinline{open} code

\begin{center}\begin{tabular}{rcl}
\begin{lstlisting}
open struct type t = e end
\end{lstlisting}
\end{tabular}\end{center}

\noindent
i.e.~each occurrence of \lstinline{t} in the remainder of the
signature would be replaced by \lstinline{e}.

This design supports the signature use cases in
Section~\ref{section:signature-use-cases}.
For example, using local aliases, the module on the left below can be
given the signature on the right:

\begin{center}
\begin{tabular}{cp{0.5in}c}
\begin{lstlisting}
type t = T1
module M = struct
  type t = T2
  let f T1 = T2
end
\end{lstlisting}&&
\begin{lstlisting}
type t = T1
type t' := t
module M : sig
  type t = T2
  val f : t' -> t
end
\end{lstlisting}
\end{tabular}
\end{center}

Furthermore, like destructive substitution itself, this extended
design is restricted to module and type aliases, and so the syntactic
concerns with \lstinline{private} (Section~\ref{section:private}) do
not arise.

\section{Status}
\label{section:full-status}

A variant of the design proposed in this article was discussed at the
Caml developers meeting and accepted for inclusion into OCaml 4.08.
The subsequent GitHub pull request and further discussion may be found
at the following URL:

\begin{quotation}
\url{https://github.com/ocaml/ocaml/pull/1506}
\end{quotation}

The design for extended \lstinline{open} in structures has been
incorporated directly, and so the use cases of
Section~\ref{section:use-cases} can be used as written in OCaml 4.08.

Furthermore, \lstinline{open} in signatures has been extended beyond
simple paths to support functor application (e.g.~\lstinline{open F(X)}), and it is anticipated that it will eventually be further
extended to support transparent ascription
(Section~\ref{section:restricted-open}) and structures containing only
aliases (e.g.~\lstinline{open struct type t = int end}).

However, OCaml 4.08 does not support arbitrary module expressions
as the arguments of \lstinline{open} in signature contexts,  so the
examples of Section~\ref{section:signature-use-cases} cannot be
written directly.
Instead, the release also adds support for signature-local bindings
(Section~\ref{section:signature-local-bindings}), which covers
those use cases.

\section*{Acknowledgments}
\label{section:acknowledgements}

We thank Leo White and the OCaml'17 workshop and post-proceedings
reviewers for comments and suggestions, and Alain Frisch and Thomas
Refis for help with the implementation.

\bibliographystyle{eptcs}
\bibliography{open}
\end{document}